\documentclass[preprint,showpacs,showkeys,preprintnumbers,amsmath,amssymb,prb]{revtex4}
\usepackage{graphicx,epsfig}
\usepackage{amssymb}
\usepackage{bbm}
\usepackage{subfigure}
\usepackage{sidecap}
\usepackage{mathtools}
\usepackage{relsize}

\usepackage{color}

\begin{document}

\title{Quantum criticality near the Stoner transition in a two-dot
with spin-orbit coupling}

\author{\bf Oleksandr Zelyak}
  \email{zelyak@pa.uky.edu}
\author{\bf Ganpathy Murthy}
  \email{murthy@pa.uky.edu}
\affiliation{
Department of Physics and Astronomy,
University of Kentucky, Lexington, Kentucky 40506, USA
}

\date{\today}

\begin{abstract}
We study a system of two tunnel-coupled quantum dots, with the first
dot containing interacting electrons (described by the Universal
Hamiltonian) not subject to spin-orbit coupling, whereas the second
contains non-interacting electrons subject to spin-orbit coupling.  We
focus on describing the behavior of the system near the Stoner
transition. Close to the critical point quantum fluctuations become
important and the system enters a quantum critical regime. The
large-$N$ approximation allows us to calculate physical quantitites
reliably even in this strongly fluctuating regime. In particular, we
find a scaling function to describe the crossover of the quasiparticle
decay rate between the renormalized Fermi liquid regime and the
quantum critical regime.
\end{abstract}



\maketitle

%
%
%
      \section{Introduction}
%
%

   The transport of electrons through mesoscopic systems at low 
temperatures is a coherent process. The manifestations of coherent
electronic motion are weak localization, Aharonov-Bohm oscillations,
persistent current, etc.\cite{Akkermans_book_07}

Though many mesoscopic effects can be explained in the framework of
non-interacting electrons, there is a vast amount of evidence showing
that collective effects of the electron spin are important in
predicting the behavior of the system. Apart from its fundamental
interest, studying interaction-related effects on the electronic spin
is important from a technological point of view.  

The spin of the electron couples to the external magnetic field and to
the orbital degrees of freedom. This spin-orbit coupling (SO) is
caused by a non-zero electric field in the laboratory reference frame
that is transformed into a magnetic field in the electron's rest
frame.  In bulk systems the SO coupling results from the absence of
inversion symmetry in the crystalline lattice\cite{Dresselhaus_1955}
(Dresselhaus term). In finite size systems, such as metallic grains or
semiconductor quantum dots, an additional contribution to SO coupling
comes from the structure inversion asymmetry\cite{Bychkov_1984}
(Rashba term), the simplest example of which is a two-dimensional
electron gas (2DEG) confined to an interface, in which the confining
electric field perpendicular to the 2DEG is the source of the SO
coupling.

In diffusive and ballistic/chaotic mesoscopic systems the kinetic term
in the full Hamiltonian is well described by Random Matrix Theory
\cite{Mehta_04, stockmann:book} (RMT). RMT has been very successful in
describing the ensemble averages of one-particle spectral correlations
as well as correlations of eigenfunctions.

RMT describes the universal zero-dimensional limit in a mesoscopic
system. Its regime of validity is when all time scales (the spin-orbit
relaxation time $\tau_{SO}$ and the inverse mean level spacing
$\delta^{-1}$) are much larger than ergodic time
$\tau_{erg}=\hbar/E_T$. Alternatively, all relevant energy scales
should be smaller that Thouless energy $E_T$ (for a diffusive dot of
linear scale $L$, $E_T\simeq \hbar D/L^2$, where $D$ is the diffusion
constant, while for a ballistic/chaotic dot $E_T\simeq \hbar v_F/L$).

Even though $\tau_{SO} \gg \tau_{erg}$ defines the universal limit, to
decide if the SO coupling is important for a particular physical
process, $\tau_{SO}$ should be compared to other characteristic time
scales.\cite{Brouwer_2002,gorokhov_2004,Oreg_all_2001} As the SO
coupling is increased from zero in a noninteracting system, its
effects begin to become important for physical quantities when the
inverse SO relaxation time is comparable to the mean level spacing
$\delta \sim \tau_{SO}^{-1}$. In interacting systems near a degeneracy
point between two ground states of different spin, even much tinier SO
couplings can have order one effects.\cite{murthy-shankar-SO}

RMT systems can be classified according to the presence or absence of
time reversal (TR) and spin rotation symmetries. They fall into three
major categories described by the ``classical'' RMT ensembles
introduced by Dyson. The systems with both symmetries preserved belong
to the Gaussian Orthogonal ensemble.  Systems with broken TR symmetry
(e.g. by an external magnetic field) are described by the Gaussian Unitary
ensemble. Finally, systems with TR preserved and broken spin rotation
symmetry belong to the Gaussian Symplectic ensemble.

In the universal regime, the SO coupling has significant effect on
spectral properties of eigenvalues and eigenfunctions. One can relate
the spin-orbit scattering length $L_{SO}$ to a SO crossover energy
scale $E_{X}=E_T\big({L\over L_{SO}}\big)^2$. For energies below
$E_{X}$, the one-particle term in the Universal Hamiltonian is modeled
by a Gaussian symplectic random matrix. If one wants to describe
energies both above and below $E_{X}$, one has to use the RMT ensemble
which is in a crossover between the GOE and the GSE
classes.\cite{Mehta_04,adam:article}

The interactions in mesoscopic systems at low temperatures are
described by the Universal
Hamiltonian:\cite{Andreev_Kamenev_81,Brower_Oreg_Halper_99,
Baranger_Ulmo_Glazman_00,Kurland_Aleiner_Altshuler_00}

\begin{equation}
   H_U = \sum_{\alpha ,s} \epsilon_{\alpha} c^{\dagger}_{\alpha ,s}
   c_{\alpha ,s} + \frac{U_0}{2} \hat{N}^2 - J{\bf S}^2
 + \lambda T^{\dagger} T,
\label{eq:H_U}
\end{equation}

where $\hat{N}$ is the total particle number, ${\bf S}$ is the total
spin, and $T=\sum c_{\beta ,\downarrow} c_{\beta ,\uparrow}$. The
Universal Hamiltonian contains a charging energy $U$ (direct channel),
a Stoner exchange energy $J$ (spin channel) and a reduced
superconducting interaction $\lambda$ (Cooper channel). A
renormalization group (RG) analysis
reveals\cite{murthy-mathur,murthy-shankar1} that this is the
low-energy effective theory for weak coupling, although other
effective theories and other ground states can be accessed for strong
coupling.\cite{murthy-mathur-shankar}

For small normal metallic grains and non-superconducting quantum dots
with a fixed number of particles the exchange interaction is the main
contribution to electron-electron interactions.  The short range part
of electron-electron interactions causes the ferromagnetic Stoner instability
at large values of exchange energy $J$.

In the absence of SO coupling the total spin of the system 
${\bf S}^2$ and its z-projection $S_z$ commute with kinetic energy term 
and are good quantum numbers. Typically for metallic grains the exchange
constant $J\lesssim \delta$.

For weak exchange interaction $J \ll \delta$ the spin of the ground
state for odd number of electrons is $1/2$. As $J$ gets larger, there
is a non-zero probability to obtain a ground state with $S>1/2$. This
happens when the cost in orbital energy to promote an electron to the
next level is less than the energy gain due to the exchange
interaction. As J approaches $\delta$, the total spin of the system
grows\cite{Kurland_Aleiner_Altshuler_00}, and at $J=\delta$ the system 
undergoes a phase transition (the Stoner transition) into a ''bulk'' 
ferromagnetic state. For $J \ge\delta$ the magnetization of the system 
is proportional to the number of electrons $N$.

In the presence of SO coupling the total spin does not commute with
full Hamiltonian $[H,{\bf S}^2] \ne 0$. While the dominant effect of
the electron-electron interaction is to organize the states according
to total spin ${\bf S}$, the SO term produces matrix elements between
states of different spin, which randomizes spin, and also leads to
sample-to-sample fluctuations of the matrix elements of the
electron-electron interaction\cite{adam:article} and the suppression
of the exchange interaction.\cite{gorokhov_2004,Alhassid_Rupp_03:condmat}

When $J \gg \gamma_{SO}$, the SO coupling is simply ignored (unless
one is near a degeneracy between ground states of different
spin.\cite{murthy-shankar-SO}) In the opposite limit $J \ll \delta$ and
$J \ll \gamma_{SO}$ electron-electron interactions are suppressed and
expectation value of total spin in ground state $\langle S \rangle <
1/2$.  The interesting regime is when $\gamma_{SO}\sim J \lesssim
\delta$. In this case the  exchange interaction is not completely
suppressed, and the fluctuations of total spin are comparable to its
expectation value. This regime is driven by the combined effect of
spin-orbit scattering and electron-electron interactions.

We study the regime where the system is near the Stoner instability $J
\rightarrow \delta^{-}$. If SO coupling is absent, there are no
quantum fluctuations of the spin, and one obtains a sequence of
metamagnetic transitions with the true Stoner transition being the
accumulation point.\cite{Kurland_Aleiner_Altshuler_00} In the presence 
of SO coupling, at low energies, the behavior of the system is dominated 
by quantum critical fluctuations leading to the formation of a quantum 
critical regime (QCR).\cite{Sachdev:book,murthy:article}

Imagine that one is at some $J<\delta$, but that
$1-J/\delta\ll1$. Even close to the transition one can think of two
different regimes of energy separated by a many-body crossover scale
$E_{QCX}$, which will turn out to be simply related to the
single-particle RMT crossover scale $E_X$.\cite{murthy:article} For
$\omega\ll E_{QCX}$ the system behaves as though it were a
renormalized Fermi liquid, with altered Fermi liquid parameters and a
quasiparticle decay rate going as $\omega^2$.\;\cite{Altshuler_Gefen_1997,Sivan_Imry_1994}
On the other hand, for
$E_{QCX}\lesssim \omega$, the behavior is controlled by the quantum
critical point. The change of behavior as one increases $\omega$ is
described by a universal scaling function $F(\omega/E_{QCX})$.

The critical point and QCR are dominated by many-body quantum
fluctuations, and thus the scaling functions cannot be calculated
perturbatively. However, it turns out that as long as
$E_{QCX},\omega\gg \delta$, one can use a large-$N$ approximation with
$min\big({E_{QCX}\over\delta},{\omega\over\delta}\big)$ playing the
role of the large $N$.\cite{murthy:article} This allows us to compute
the scaling functions reliably.

From the point of view of experiment, the key point is that one can
control $E_{QCX}$, which is a many-body scale, by tuning a
single-particle crossover energy scale $E_{X}$. Thus, at a fixed value
of the parameter $J$, one can tune oneself into and out of the QCR by
tuning a single-particle knob.

As a prerequisite to describing the system near Stoner transition, we
consider the non-interacting case and calculate ensemble-averaged one
and two particle Green's functions for electrons in the first dot
coupled to the second dot in crossover between GOE and GSE
ensembles. The one particle Green's function is unchanged by
crossover, though it is modified by interdot coupling. The two
particle Green's function is the sum of the contributions due to
diffuson\cite{efetov:book} mode and Cooperon modes. Bot contributions
depend on the ratios of crossover parameter $E_{X_2}$, interdot
coupling parameter $E_U$, and measurement energy $\omega$.

It may seem counterintuitive that one can use {\it non-interacting}
wavefunction averages to describe the behavior of a system with strong
many-body fluctuations\cite{adam:article,Adam_Brouwer_Sharma_03,
Alhassid_Rupp_03:condmat,murthy:article}, but this goes hand in hand
with the use of the large-$N$ approximation. This is because there is
no wavefunction renormalization to leading order in the the large-$N$
approximation.

%
%
%
      \section{Model Definition}
%
%

\begin{figure*}
  \centerline{
    \mbox{\includegraphics[width=4.5in]{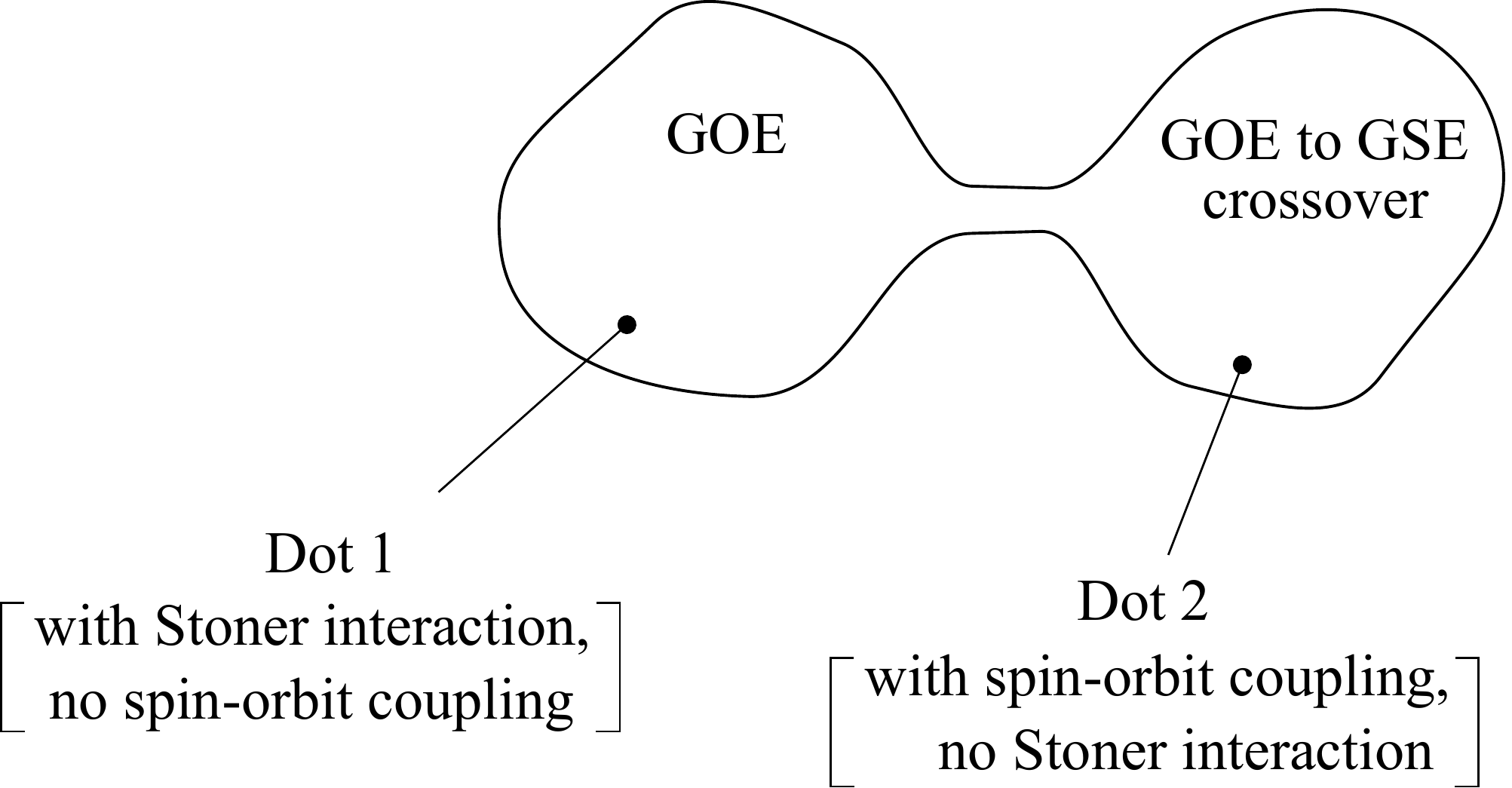}}
  }
  \caption{Two coupled quantum dots in GOE-to-GSE crossover.}
  \label{CoupledDots}
  \end{figure*}

We consider a system of two quantum dots (metallic grains) coupled to
each other by tunneling (see Fig.\;\ref{CoupledDots}). The motion of
electrons can be either diffusive or ballistic/chaotic: in either case
the single-particle energies and wavefunctions are controlled by RMT,
which is all that we require.


For the non-interacting system the ensemble-averaged spectral and
eigenvector correlations can be computed by
RMT.\cite{Mehta_04,stockmann:book} The first dot belongs to the GOE,
since it has no spin-orbit coupling. The second dot has weak spin
orbit coupling that drives it into the GOE$\rightarrow$GSE crossover,
characterized by a crossover scale $E_{X_2}$. We also assume that
there is a Stoner exchange interaction in the first dot. No
interactions are present in the second dot. The tunneling between the
dots gives rise to another crossover scale $E_U$, where $E_U/\delta$
is the dimensionless conductace between the two dots.

\begin{figure*}
  \centerline{
    \mbox{\includegraphics[width=3.5in]{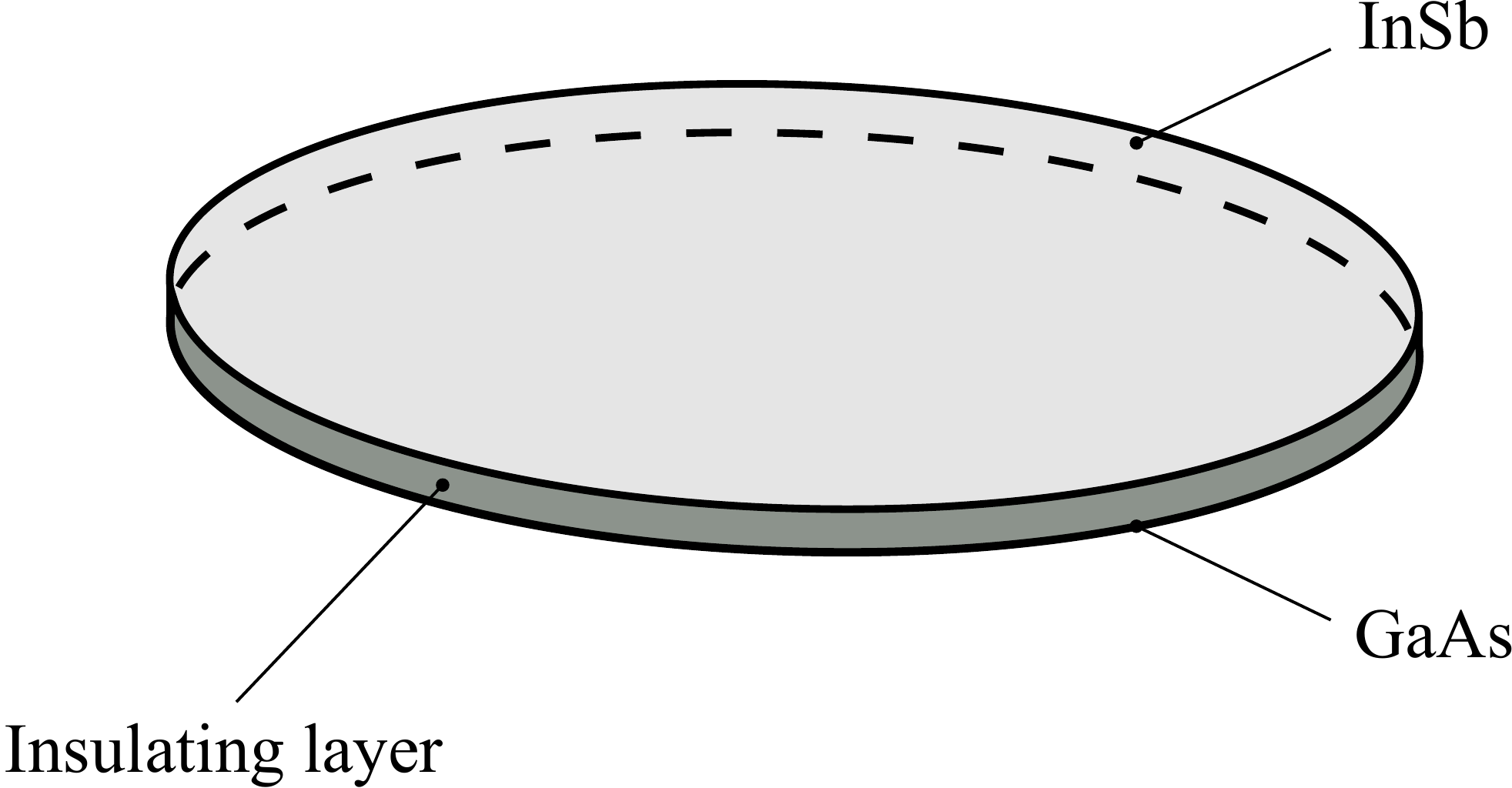}}
  }
  \caption{(Color online) The vertical arrangement of the dots allows us to get rid of
                        the charging energy.}
  \label{VerticalDots}
  \end{figure*}

In Fig.\;\ref{VerticalDots} one can see a more realistic picture of
the system.  In an experimental setup the lower dot could be made of
{\it GaAs} (with significant exchange interaction but tiny SO
coupling), while the upper dot could be made of {\it InSb} (with large
spin-orbit coupling). The choice of the vertically coupled geometry
will be discussed below.


In the low-energy limit interactions are described by Universal 
Hamiltonian.\cite{Andreev_Kamenev_81,Brower_Oreg_Halper_99,
Baranger_Ulmo_Glazman_00,Kurland_Aleiner_Altshuler_00} \eqref{eq:H_U}
For our system $H_U$ has the form:

\begin{equation}
  \begin{split}
  H &= \sum_{i_0 j_0 s} H^{(1)}_{i_0 j_0} c^{\dagger}_{i_0 ,s} c_{j_0 ,s}
    - J{\bf S}^2
 + \sum_{\mu_{0}\nu_{0} s} H^{(2)}_{\mu_{0}\nu_{0}} c^{\dagger}_{\mu_{0} ,s} 
   c_{\nu_{0} ,s}
    + \sum_{i_0 \mu_0 s} V_{i_0 \mu_0} (c^{\dagger}_{i_0 ,s} c_{\mu_0 ,s}
    + H.c)\\
    &= \sum_{\mu \tau} \epsilon_{\mu} c^{\dagger}_{\mu ,\tau}
       c_{\mu ,\tau} - J{\bf S}^2,
  \end{split}
\label{eq:our.model_H_U}
\end{equation}
  where $H^{(2)}$ contains the effect of spin orbit coupling in the
second dot. In \eqref{eq:our.model_H_U} we have omitted
the superconducting term as irrelevant to our model. We also choose a
vertically coupled geometry for our system to minimize the change in
charging energy when the electron hopes from one dot to
another.\cite{Zelyak_Murthy_2007} If this energy is smaller than all
other relevant scales, then the charging term can be omitted (or
absorbed into the chemical potential) since the total number of
electrons in the two-dot system remains unchanged.

The $(\mu,\tau)$ label the basis of the two coupled quantum dots without
interaction, that is, it is the set of eigenstates of
$H^{(1)}+H^{(2)}+V$.  Here $\mu$ is the orbital quantum number and
$\tau$ is a twofold degenerate Kramers index.  In this basis the $a$-th
component of total spin reads\cite{Alhassid_Rupp_03:condmat}

\begin{equation}
  S^a = \sum_{i_0ss'}c^{\dagger}_{i_0s} \frac{\sigma^a_{s s^{'}}}{2}c_{i_0s'}=\sum_{\mu\tau ,\nu\tau^{'}} (M^a)^{\mu\tau}_{\nu\tau^{'}}
        c^{\dagger}_{\mu\tau} c_{\nu\tau^{'}},
  \label{eq:Sa}
\end{equation}
  where $i_0$ is an Orthogonal basis in the first dot alone (it could
  be the eigenbasis of $H^{(1)}$ but it does not have to be), and the
  matrix element $M^a$ is defined as:

\begin{equation}
  (M^a)^{\mu\tau}_{\nu\tau^{'}} = \sum_{i_0,s s^{'}} \psi^{\ast}_{\mu\tau}(i_0,s)
                    \frac{\sigma^a_{s s^{'}}}{2} \psi_{\nu\tau^{'}}(i_0,s^{'}).
  \label{eq:Ma}
\end{equation}

We reiterate that the first summation in Eq.\eqref{eq:Sa} and the
summation in Eq.\eqref{eq:Ma} is over an Orthogonal basis in the first
dot alone, while the second summation in Eq.\eqref{eq:Sa} is over the
eigenbasis of the total non-interacting Hamiltonian
$H^{(1)}+H^{(2)}+V$. Also, $\sigma^a_{s s^{'}}$ are the Pauli
matrices, $\psi_{\mu\tau}(i_0,s)$ is the wave function of the state
$\mu,\tau$ in the first dot.

We use Eq.\eqref{eq:our.model_H_U} to calculate the partition function 
$Z=Tr[\exp(-\beta H)]$, using the imaginary time path integral formalism:

\begin{equation}
  Z = Tr(e^{-\beta H}) \Rightarrow \ \ 
  Z = \int \prod_{\mu\tau} \mathcal{D} \bar{c}_{\mu\tau} \mathcal{D} c_{\mu\tau} 
      \mathcal{D} {\bf h}\ e^{-\int_0^{\beta}\mathcal{L}dt},
  \label{eq:Z_sum}
\end{equation}
  where the Euclidean Lagrangian is:

\begin{equation}
  \mathcal{L} = \frac{\vert {\bf h} \vert^2}{4J}
              + \sum_{\mu\tau} \bar{c}_{\mu\tau} (\partial_t + \epsilon_{\mu}) 
                c_{\mu\tau} - {\bf h\cdot S}.
  \label{eq:Lagrangian}
\end{equation}

In Eq.\eqref{eq:Z_sum} we used the Hubbard-Stratonovich transformation
to decouple the interaction at the expense of introducing an
additional bosonic field {\bf h} representing the order parameter. The
$c_{\mu\tau}$ and $\bar{c}_{\mu\tau}$ are Grassmann variables.

After switching to the Fourier representation the fermionic fields
$c,\bar{c}$ are integrated out. The resulting action for {\bf h} is
expanded to second order to obtain:

\begin{equation}
  S_{eff} \approx \frac{1}{4\beta\delta_1}\sum_{n,a} \vert h^a(i\omega_n)\vert^2 
  \Big[\frac{1}{\tilde{J}} - f_n(\beta,E_{X_2},E_U) \Big]
\end{equation}

\begin{equation}
  f_n(\beta,i\omega_n) 
    = -2\delta_1\sum_{\mu\tau,\nu\tau^{'}}{\vert M^a \vert^2}^{\,\mu\tau}_{\,\nu\tau^{'}}\ 
    \frac{N_F(\epsilon_{\mu}) - N_F(\epsilon_{\nu})}{\epsilon_{\mu} - \epsilon_{\nu} - i\omega_n}
  \label{eq:f_n}
\end{equation}
  where $\omega_n = 2\pi n/\beta$, $\tilde{J} = J/\delta_1$ is a
dimensionless exchange constant, $N_F(\epsilon_{\mu})$ is the
Fermi-Dirac occupation of the state $\mu$, and $\delta_1$ is a mean
level spacing for the first dot.

Deep into the crossover $E_U,E_{X_2}\gg\delta_1$, we replace ${\vert
M^a \vert^2}^{\,\mu\tau}_{\,\nu\tau^{'}}$ by its RMT ensemble
average. This is justified because in the limit when
${E_{X_2}\over\delta},{E_U\over\delta}\to\infty$ the spectral average
on a single sample is the same as the ensemble average. The
corrections to this vanish in the large-$N$ limit. This is one of the
ways in which we use the large-$N$ approximation.

The relevant four wavefunction correlator hidden in $\vert M^a
\vert^2\rangle$ is calculated in Appendix \ref{apnx:B}.  We also
replace the summation over energy eigenstates by energy
integrations. Assuming a constant density of states we obtain

\begin{equation}
 f_n(\beta,E_{X_2},E_U) = \frac{E_U}{E^2_2 - E^2_1} 
\bigg[\frac{E^2_{X_2}+E_{X_2}E_U-E^2_1}{E_1+\vert \omega_n\vert} -
     \frac{E^2_{X_2}+E_{X_2}E_U-E^2_2}{E_2+\vert \omega_n\vert}\bigg]
\end{equation}
   where the interdot tunneling energy scale $E_U$, the SO crossover
energy scale $E_{X_{2}}$ in the second dot, and the energies $E_{1,2}$
(which are functions of $E_U$ and $E_{X_2}$) are defined in Appendices
\ref{apnx:B},\ref{apnx:C}.

  The instability point is obtained by setting $f_0(\beta,E_{X_2},E_U)
= \tilde{J}^{-1}$.  For the coupled-dot system the quantum phase
transition takes place at $\tilde{J} = 1$, or $J = \delta_1$, the same
result as for one uncoupled dot independent of the crossover energy
scales.

  We investigate the limit when $E_U \ll E_{X_2}$. In this limit $E_U$
is the only relevant parameter that controls both the coupling between
dots and the degree to which spin rotation symmetry is spoiled in the
first dot. In this limit the scaling function becomes $f_n = E_U/(E_U
+ \vert \omega_n \vert)$.

  Close to the transition the smallness of $1-\tilde{J}$ allows us to
introduce a new scaling function $F_n$ that describes the interacting
system near Stoner transition. The effective action now becomes:

\begin{equation}
  S_{eff} = \frac{1}{4\delta_1\beta} \sum_{n,a} \vert h^a(i\omega_n) \vert^2 F_n
\end{equation}

\begin{equation}
  F_n = \frac{E_{QCX}}{\tilde{J}E_U} \Big(1 + \frac{\vert \omega_n \vert}{E_{QCX}}\Big)
  \label{eq:F_n}
\end{equation}

  The scaling function $F_n$ in \eqref{eq:F_n} describes how a
physical quantity behaves when one goes from the renormalized Fermi
liquid to the quantum critical regime.  The new characteristic energy
scale $E_{QCX} = E_U (1-\tilde{J})$ can be used to tune the system
into the QCR. By changing the single particle parameter $E_U$ in
$E_{QCX}$ one can access the QCR governed by interactions.

\begin{figure*}
  \centerline{
    \mbox{\includegraphics[width=3.5in]{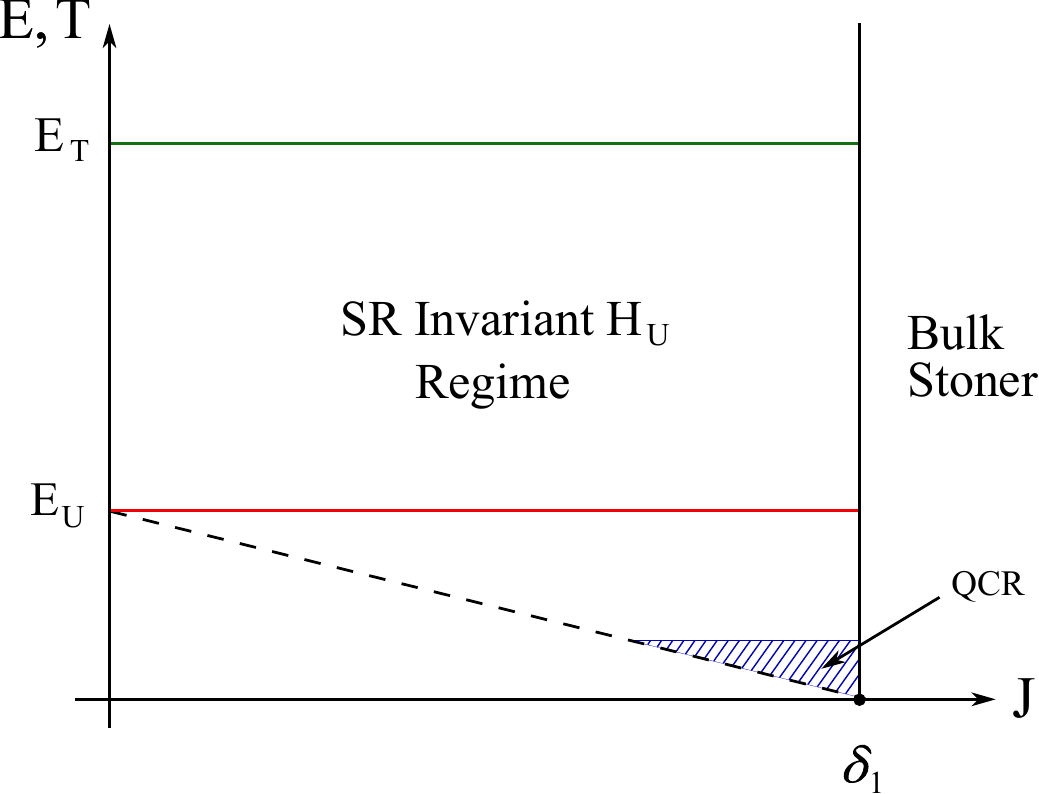}}
  }
  \caption{(Color online) Phase diagram in the $E$ vs. $J$ showing different regimes.}
  \label{QCR_diagram}
  \end{figure*}

  The Fig.\;\ref{QCR_diagram} shows the phase diagram in $(\omega ,T)$
vs. $J$ coordinates. For $J > \delta_1$, the system is in a ``bulk''
Stoner phase where the magnetization is proportional to the volume of
the system. When $J<\delta_1$ and the measurement energy $\omega$
satisfies the inequality $E_U\ll\omega<E_T$ one enters an approximate
spin-rotation-invariant universal regime described by Universal
Hamiltonian $H_U$. Here the total spin of the system is
(approximately) a good quantum number. Lowering the energy to $\omega
\sim E_U$ brings us to regime where the system starts seeing
spin-orbit coupling and spin fluctuations become important. Below the
line $E = E_U(1-\tilde{J})$ is the renormalized Fermi liquid
regime. Above this line, away from the Stoner instability point, is
the non-universal regime that depends on many parameters. Close to the
instability point $J=\delta_1$ is a Quantum Critical Regime controlled
by a single parameter $E_{QCX}$. One can change the single-particle
parameter $E_U$ in $E_{QCX}$ to access the many-body regime.


  We proceed to calculate the quasiparticle decay rate near the
critical point.  Since the particle decays by interacting with quantum
fluctuations of collective spin, the decay rate can be obtained from
the spin-spin correlation function $\langle S^a(t)S^b(t^{'}) \rangle$
which, in turn, can be measured by NMR or EPR.  In Fourier space the
$S^aS^b$ correlator can be expressed through the bosonic field $h^a$
as follows (see Appendix \ref{apnx:C} for more details)

\begin{equation}
  \langle S^a(i\omega_n) S^b(-i\omega_n) \rangle = -\frac{\delta_{ab}}{2J}
  + \frac{1}{4J^2} \langle h^a(i\omega_n) h^b(-i\omega_n) \rangle.
\end{equation}

Calculating the $\langle h^ah^b \rangle$ correlator

\begin{equation}
  \langle h^a(i\omega_n) h^b(-i\omega_n) \rangle = Z^{-1}
  \int \mathcal{D} {\bf h} h^a(i\omega_n) h^b(-i\omega_n) e^{-S_{eff}}
  = \delta_{ab}\frac{4JE_U}{E_{QCX}\big( 1+\frac{\vert \omega_n \vert}{E_{QCX}} \big)}
  \label{eq:hh_expression}
\end{equation}
  one obtains the following spin-spin correlation function

\begin{equation}
  \langle S^a(i\omega_n) S^b(-i\omega_n) \rangle = -\frac{\delta_{ab}}{2J}
  \Big[ 1 - \frac{2E_U}{E_{QCX}\big( 1+\frac{\vert \omega_n \vert}{E_{QCX}} \big)}\Big]
  \label{eq:SS_real_time}
\end{equation}

\begin{figure*}
  \centerline{
    \mbox{\includegraphics[width=3.5in]{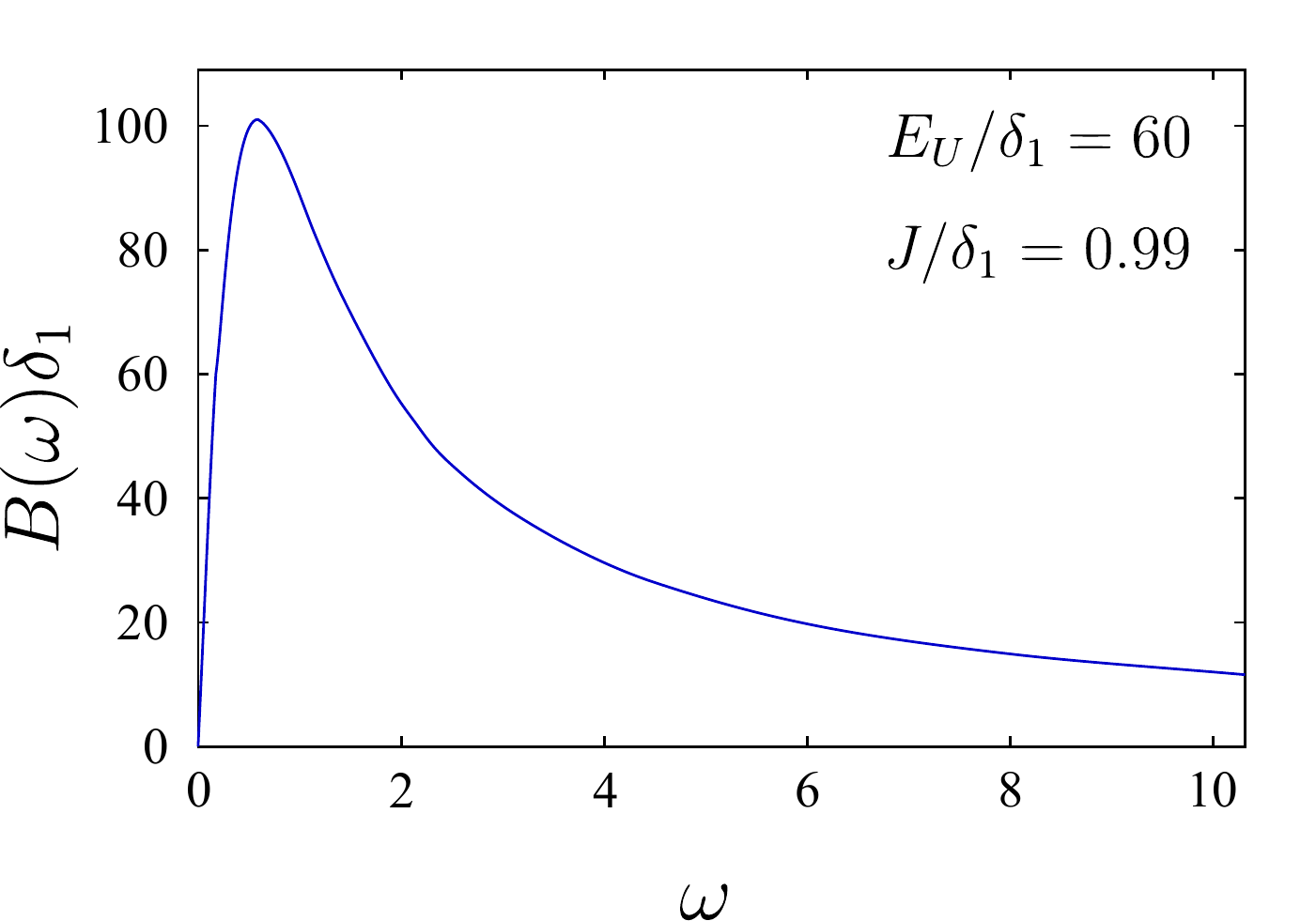}}
  }
  \caption{{(Color online) Spectral function for spin-spin excitations.}}
  \label{Spectral_function}
  \end{figure*}

  Switching back to the real time formalism $(i\omega_n\to\omega + i\eta, \eta \to0^+)$ 
in Eq.\eqref{eq:SS_real_time} one obtains the spectral function of
spin excitations

\begin{equation}
  B(\omega) = -2\Im \Big[ S^a(\omega) S^b(-\omega)\Big] = \delta_{ab}
  \frac{2E_U}{J} \frac{\omega}{\omega^2 + E^2_{QCX}}.
  \label{eq:B_spectr_fn}
\end{equation}
  The graph of the spectral function \eqref{eq:B_spectr_fn} is shown 
 on Fig.\;\ref{Spectral_function}.


  The decay rate of quasiparticles is found by estimating the
lowest-order interacting self-energy diagram
\includegraphics[width=1.0cm]{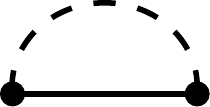} with interaction $V$:

\begin{equation}
  V = - {\bf hS} = -\sum_{\mu\tau ,\nu\tau^{'}} {\bf h}{\bf M}^{\,\mu\tau}_{\,\nu\tau^{'}}
  \bar{c}_{\mu\tau} c_{\nu\tau^{'}}
\end{equation}

  The imaginary part of self-energy $\Sigma^{(1)}$ is evaluated to

\begin{equation}
  \Im \Sigma^{(1)} = \delta_{ab}\frac{J}{16\pi}\ln
  \bigg[ \frac{E_{QCX}^2(\epsilon^2 + E_1^2)}{E_1^2(\epsilon^2 + E^2_{QCX})}\bigg].
\end{equation}

\begin{figure*}
  \centerline{
    \mbox{\includegraphics[width=3.3in]{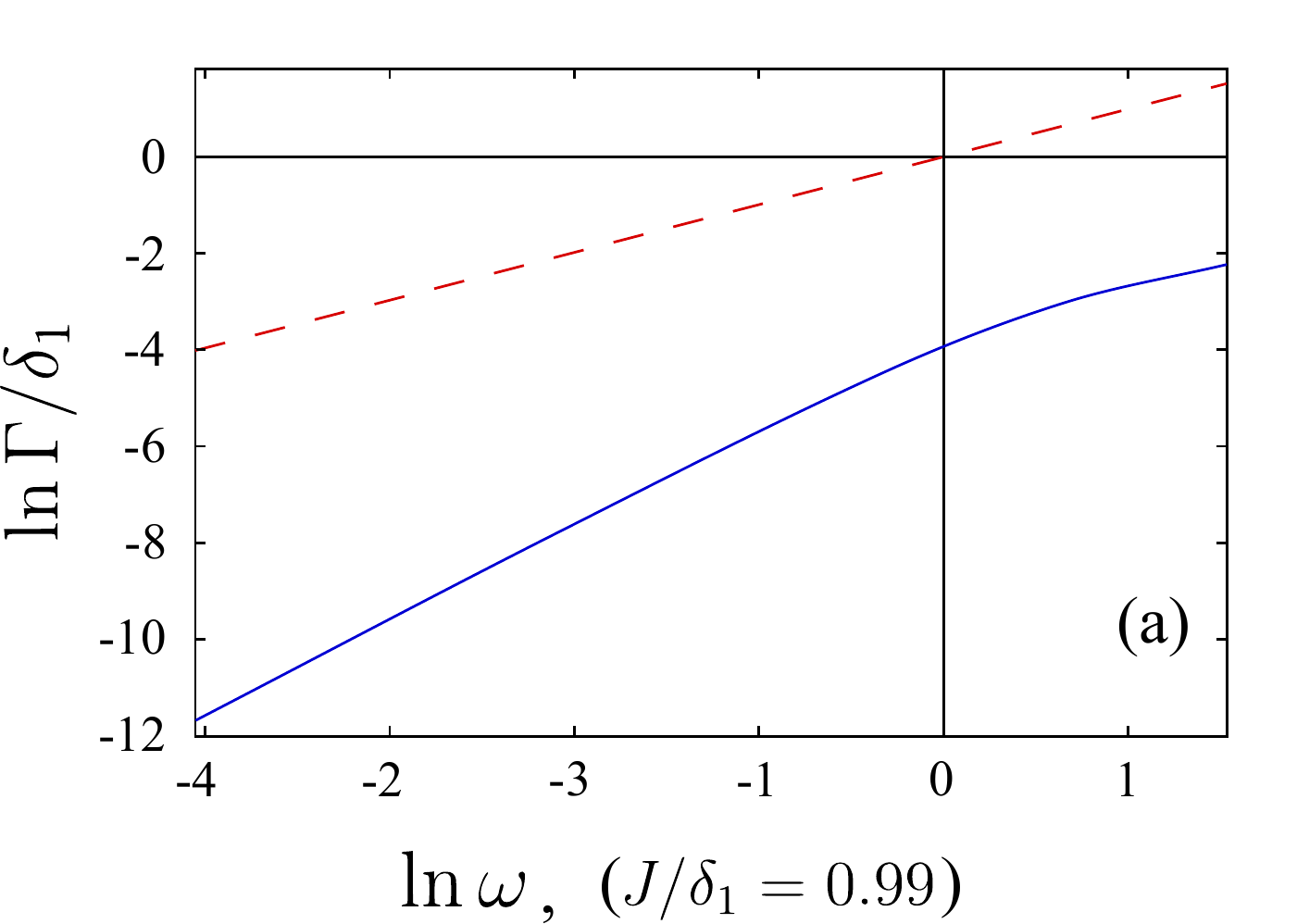}}
    \mbox{\includegraphics[width=3.3in]{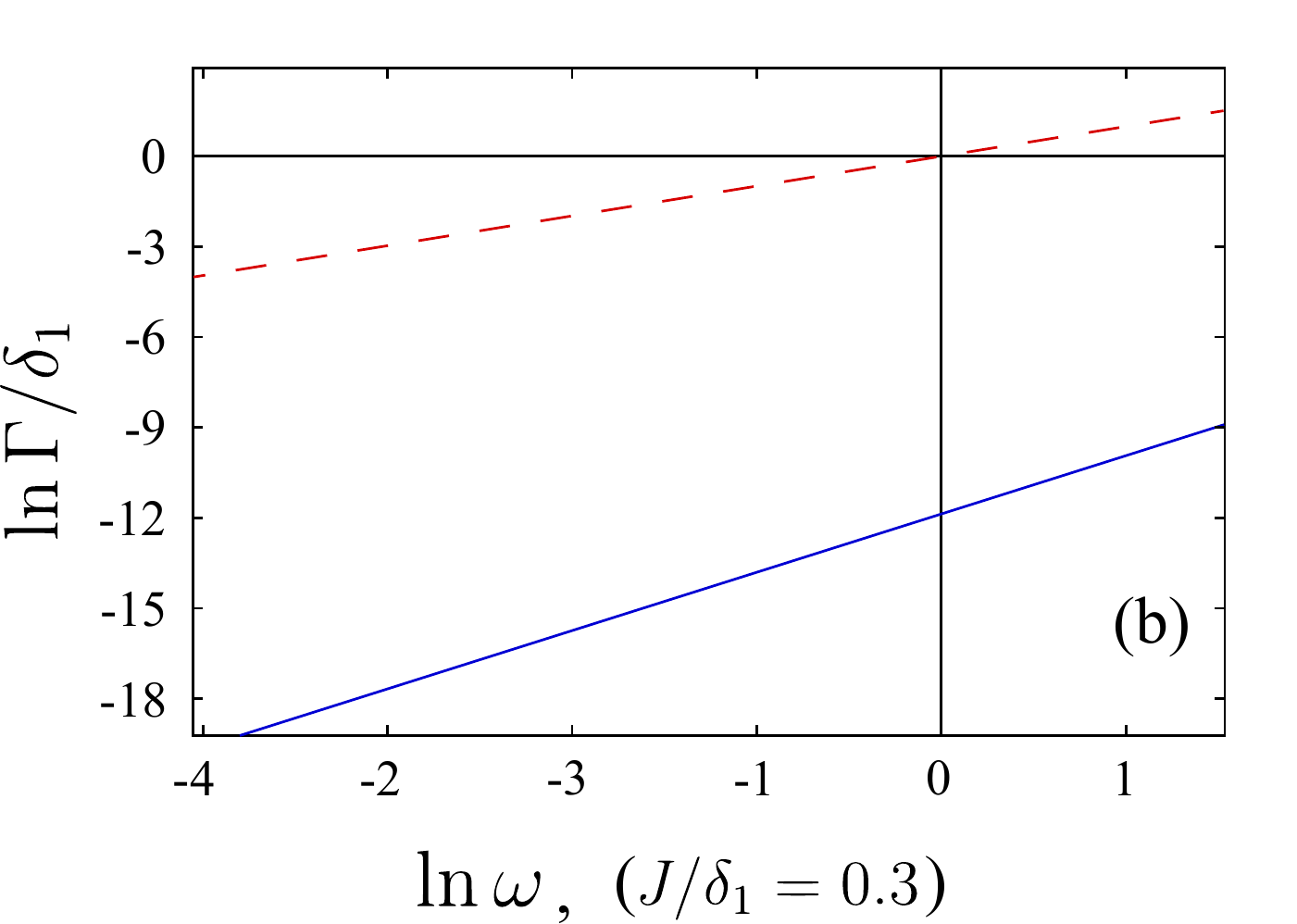}}
  }
  \centerline{
    \mbox{\includegraphics[width=3.3in]{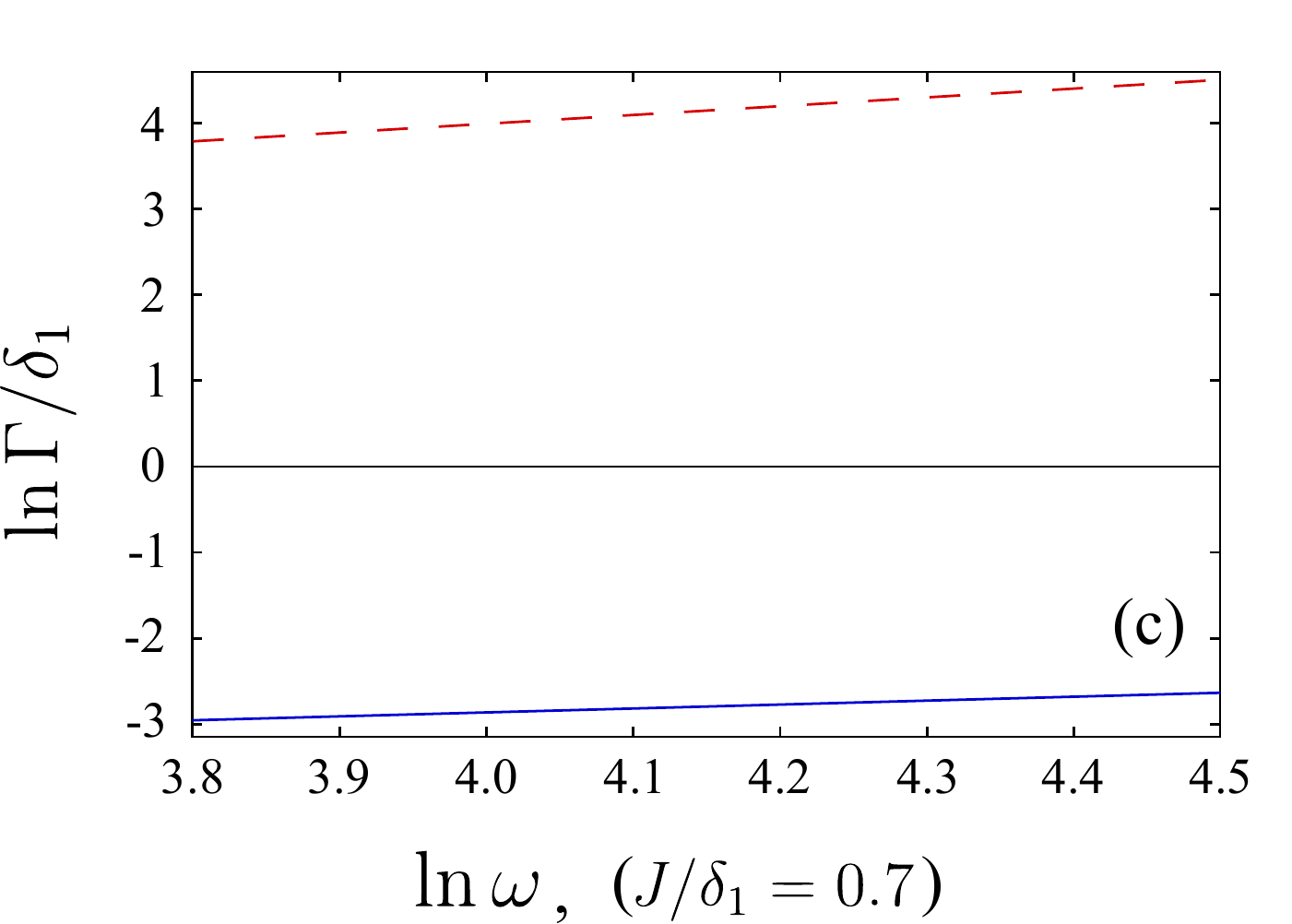}}
    \mbox{\includegraphics[width=3.3in]{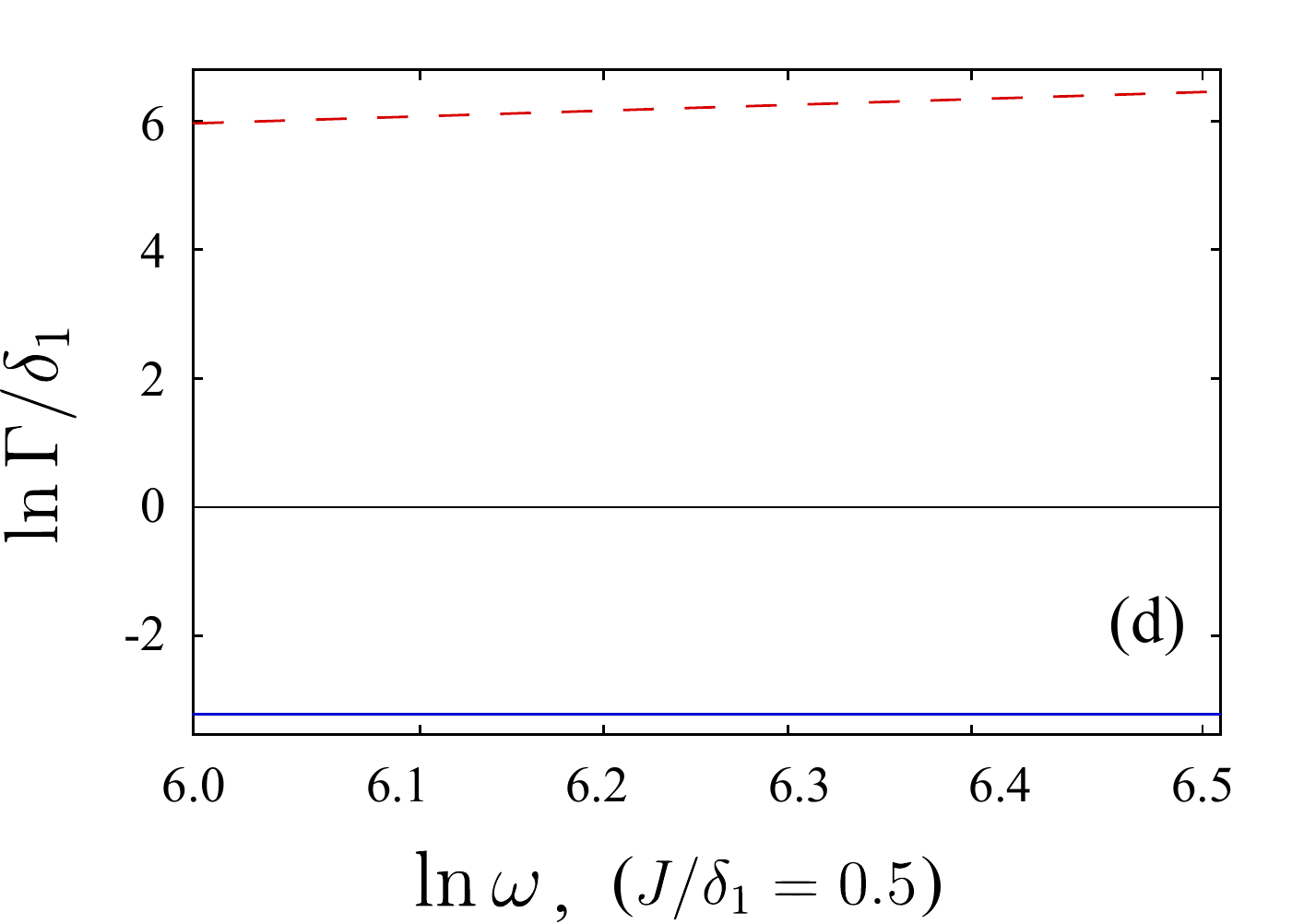}}
  }
  \caption{(Color online) Quasiparticle decay rate (solid line) on a logarithmic scale for different 
           regimes. The dashed line represents $\ln\omega$. Panel (a) shows the decay rate in quantum
           chritical regime.  The decay rate in the regime of renormalized Fermi liquid is in
           panel (b). Panels (c) and (d) show the decay rate in non-universal and Universal regimes
           respectively.}
  \label{Decay_rate}
  \end{figure*}

  The decay rate $\Gamma$ for various regimes is plotted on logarithmic scale in Fig.\;\ref{Decay_rate}.


%
%
%
      \section{Conclusion}
%
%

In this paper we have studied a system of two tunnel-coupled quantum
dots (small normal metallic grains) near the Stoner transition of the
first dot. The first dot has interacting electrons but no spin-orbit
coupling, while the other has spin-orbit coupling, but no
interactions. The two single-particle crossover energies are $E_U$,
which measures the tunneling strength between the dots, and $E_{X_2}$
which measures the spin-orbit crossover scale in the second
dot. Electrons tunneling between the dots carry information about
spin-rotation invariance breaking to the first dot, and produce
quantum fluctuations of the first dot's spin.

Our focus is on the regime near the Stoner transition when the
exchange interaction $J$ is comparable to the mean level spacing
$\delta_1$ in the first dot. This regime is characterized by quantum
critical fluctuations rising from the interplay between the spin-orbit
and interaction parts of the Hamiltonian. For this Quantum Critical
Regime we derived the scaling function describing the behavior of
system observable near instability point $J = \delta_1$ as a function
of the measurement energy $\omega$. The scaling function itself is
dependent on a single dimensionless ratio $\omega/E_{QCX}$, as opposed
to other parameter regimes where a physical property can depend
separately on all the energy scales $E_U, E_{X_2}, \omega, \delta$.

As an illustrative example we compute the scaling form of the
quasiparticle decay rate, which can be measured by nonlinear
conductance measurements. It has a Fermi liquid-like form for
$\omega\ll E_{QCX}$, with the decay rate going as $\big({\omega\over
E_{QCX}}\big)^2$. However, for $\omega\gg E_{QCX}$ it goes as
$\log(\omega/E_{QCX})$. 

One of the main conceptual points we wish to make is that there is an
intimate relation between the single-particle crossover energies and
the many-body quantum critical crossover scale. In the simplest case
$E_U\ll E_{X_2}$ this relation is $E_{QCX}=E_U(1-J/\delta)$. Access to
the quantum critical regime can be tuned by changing a single-particle
parameter.

An important open question is the effect of quantum criticality on
Coumlomb Blockade, that is, how are the distributions of the peak
positions, heights, and widths affected by quantum criticality. We
hope to explore this and other issues in future work.

\acknowledgements The authors are grateful to the National Science
Foundation for partial support under DMR-0703992. GM also wishes to
thank the Aspen Center for Physics where some of the work was carried
out.



\appendix


\section{GOE to GSE crossover in the system of two coupled dots }\label{apnx:A}

  In this appendix we derive one and two particle Green's functions for
two coupled dots in crossover between GOE and GSE ensembles. To reduce 
complexity we consider less general (but relevant to our system) situation when only second dot
is in GOE to GSE crossover. The first dot and the hopping bridge belong to GOE ensemble. 
The generalization where all parts of the system are in crossover can be worked out without
difficulty.

  The derivation will be rather sketchy for the full derivation in case of GOE to GUE crossover 
can be found in Ref.[\cite{Zelyak_Murthy_2007}].

  The Hamiltonian (kinetic part) of two coupled dots is:

\begin{equation}
  H           =
            \begin{pmatrix}
               H_{1} & V \\
               V^{\dagger} & H_{2}
            \end{pmatrix}.
\end{equation}
  where $H_{1,2}$ are the Hamiltonians for dot 1 and 2, $V$ describes coupling between two dots.
Following RMT one considers the elements of $H_{1,2}$ and $V$ as Gaussian random variables (quaternions) 
with zero mean. In the crossover between GOE and GSE Hamiltonians $H_{1,2}$ take the form:

\begin{equation}
 H_i = \frac{H_0^i\otimes I + X_i\Big[H_x^i\otimes\tau_x + H_y^i\otimes\tau_y + H_z^i\otimes\tau_z \Big] }
            {\sqrt{1 + 3X_i^2}}
  \label{eq:H_symplectic}
\end{equation}

Similarly,

\begin{equation}
 V = \frac{V^R \otimes I + \Gamma \Big[V_x^I\otimes\tau_x + V_y^I\otimes\tau_y + V_z^I\otimes\tau_z \Big] }
            {\sqrt{1 + 3\Gamma^2}}
  \label{eq:V_symplectic}
\end{equation}
  where $H_0^i$ and $H_{xyz}^i$ are real symmetric and real antisymmetric matrices. $V^R$ and $V^I_{xyz}$
are real and imaginary parts of quantum matrix $V$ (note that elements of $V^R$ and $V^I_{xyz}$ are
real numbers). The $\tau_i$ matrices are related to Pauli matrices as $\tau_k = i\sigma_k$, $k=x,y,z$.

The $X_i$ and $\Gamma$ are crossover parameters. The denominators in Eqs. \eqref{eq:H_symplectic} and 
\eqref{eq:V_symplectic} keep mean level spacing constant when $X_i$ and $\Gamma$ change.

  In calculations below we assume $X_1 = \Gamma = 0$, so the first dot and the bridge belong to GOE;
the mean level spacing $\delta_1 = \delta_2 \Leftrightarrow N_1 = N_2$.

The elements of $H_1$ and $H_2$ are independent random variables with correlations between symmetric and
antisymmetric parts

\begin{equation}
 \langle H^{s,a}_{mn} H^{s,a}_{st} \rangle = 
  \frac{N_1\delta_1^2}{\pi^2}\big( \pm \delta_{mt}\delta_{ns} + \delta_{ms}\delta_{nt}\big)
\end{equation}
  where indices s(a) stand for symmetric(antisymmetric); $N_1$ is the size of matrix $H_1$ and $\delta_1$ 
is the mean level spacing (we assume that $N_1 = N_2$, which means $\delta_1 = \delta_2$). 
Correlation between full matrix elements in crossover is

\begin{multline}
 \langle H_{m\xi_m,n\xi_n} H_{s\xi_s,t\xi_t}  \rangle = \frac{N_1\delta_1^2}{\pi^2}
 \frac{\delta_{mt}\delta_{ns}\big[ (1-X^2)\delta_{\xi_m\xi_n}\delta_{\xi_s\xi_t}
+ 2X^2 \delta_{\xi_m\xi_t}\delta_{\xi_n\xi_s} \big]}{1+3X^2}\\
\frac{N_1\delta_1^2}{\pi^2}
\frac{\delta_{ms}\delta_{nt}\big[ (1+X^2)\delta_{\xi_m\xi_n}\delta_{\xi_s\xi_t}
- 2X^2 \delta_{\xi_m\xi_t}\delta_{\xi_n\xi_s} \big]}{1+3X^2}
\end{multline}
  Here $\xi_i$ is a "spin" index that numerates elements of $\tau$ matrices.

For $V$ matrix correlations between matrix elements are

\begin{equation}
 \langle V_{nk^{'}} V_{st^{'}} \rangle = \langle V^{\dagger}_{k^{'}n} V^{\dagger}_{t^{'}s} \rangle
  = \langle V_{nk^{'}} V^{\dagger}_{t^{'}s} \rangle = \frac{N_1\delta_1^2 U}{\pi^2}
    \delta_{ns}\delta_{k^{'}t^{'}}\delta_{\xi_n\xi_{k^{'}}} \delta_{\xi_s\xi_{t^{'}}}
\end{equation}
  where primed(unprimed) indices belong to the second(first) dot; $U$ is a dimensionless parameter
controlling coupling between dots.

  One particle Green's function for coupled dots is:

\begin{equation}
 G = (E\otimes I - H)^{-1} = 
            \begin{pmatrix}
               E-H_{1} & -V \\
               -V^{\dagger} & E-H_{2}
            \end{pmatrix}^{-1}
        =   \begin{pmatrix}
               G_{11} & G_{12} \\
               G_{21} & G_{22}
            \end{pmatrix}
\end{equation}

  Following the steps in Ref.[\cite{Zelyak_Murthy_2007}] one can obtain the system 
of Dyson equations
for RMT averaged Green's functions $G_{11}$ and $G_{22}$. In large-N approximation only the 
rainbow diagrams contribute. In the limit of weak coupling the solution for $G_{11}$ is

\begin{equation}
 \langle G^R_{ab,1}  \rangle^{-1} = \delta_{ab}\delta_{\xi_a\xi_b}
 \frac{N_1\delta_1}{\pi} \Big[ \epsilon + i\sqrt{1 - \epsilon^2} \Big]
 \Big[ 1+ \frac{U}{2}\Big(1+i\frac{\epsilon}{\sqrt{1-\epsilon^2}}  \Big) \Big]
\end{equation}
  where dimensionless energy $\epsilon = \frac{\pi E}{2N_1\delta_1}$.

The two particle Green's function in the first dot can be found from the system of
Bethe-Salpeter equations.\cite{Zelyak_Murthy_2007} This system describes 
contribution of ladder and absolutely crossed diagrams.

  Expression for the full two particle Green's function is

\begin{equation}
  \langle G^R_{1,ab}(E) G^A_{1,cd}(E + \omega) \rangle = D_1 + C_1
\end{equation}

Contribution of ladder diagrams $D_1$ is

\begin{equation}
  \begin{split}
  D_1 = \delta_{ad}\delta_{bc} \frac{2\pi}{N_1^2\delta_1} \frac{1}{-i\omega}
  \bigg[ &\delta_{\xi_a\xi_b}  \delta_{\xi_c\xi_d} 
  \frac{1+i\frac{E_{X_2}+E_U}{\omega}}{\big(1+i\frac{E_U}{\omega}\big)
\big(1+i\frac{E_{X_2}+E_U}{\omega}\big) + \frac{E_U^2}{\omega^2}} \\
  +\; &\delta_{\xi_a\xi_d}  \delta_{\xi_b\xi_c}   
 \frac{-\frac{i}{2}\frac{E_{X_2}E_U^2}{\omega^3} }{ \Big[\big(1+i\frac{E_U}{\omega}\big)
  \big(1+i\frac{E_{X_2}+E_U}{\omega}\big) + \frac{E_U^2}{\omega^2} \Big] 
  \Big[ \big(1+i\frac{E_U}{\omega}\big)^2
  +\frac{E_U^2}{\omega^2}\Big] }  \bigg]
  \end{split}
\end{equation}

  Contribution of absolutely crossed diagrams is
  
\begin{equation}
  C_1 = \delta_{ac}\delta_{bd} \frac{2\pi}{N_1^2\delta_1} \frac{1}{-i\omega}
  \Big[ \delta_{\xi_a\xi_b}  \delta_{\xi_c\xi_d} \frac{\Pi^{+}_{11} + \Pi^{-}_{11}}{2} 
  + \delta_{\xi_a\xi_d}  \delta_{\xi_b\xi_c} \frac{\Pi^{+}_{11} - \Pi^{-}_{11}}{2} \Big],
\end{equation}
  where
\begin{equation*}
  \begin{split}
  \Pi^{+}_{11} &= \frac{1 + i\frac{E_{X_2} + E_U}{\omega}} {(1+i\frac{E_U}{\omega})
    (1 + i\frac{E_{X_2} + E_U}{\omega}) + \frac{E_U^2}{\omega^2}}\\
  \Pi^{-}_{11} &= \frac{1 + i\frac{E_U}{\omega}}{(1 + i\frac{E_U}{\omega})^2 + \frac{E_U^2}{\omega^2}}
  \end{split}
\end{equation*}

  Crossover energy scales $E_{X_2}$ and $E_U$ are defined as $E_{X_2} = 8X_2^2N_1\delta_1/\pi$ and
$E_U = 2U N_1\delta_1/\pi$.


\section{Correlation of four wave functions }\label{apnx:B}

  Consider the matrix element average $\langle \vert M^a \vert^2 \rangle$. More generally,

\begin{equation}
 \langle M^a_{m\tau,m^{'}\tau^{'}} M^b_{m^{'}\tau^{'},m\tau} \rangle
  = \sum_{iss^{'},i_1s_1s_1^{'}} \frac{\sigma^a_{ss^{'}}}{2} \frac{\sigma^b_{s_1s_1^{'}}}{2}
    \langle \psi^{\ast}_{m\tau}(i,s) \psi_{m^{'}\tau^{'}}(i,s^{'}) \psi^{\ast}_{m^{'}\tau^{'}}(i_1,s_1) 
                                      \psi_{m\tau}(i_1,s_1^{'}) \rangle.
  \label{eq:MM_average}
\end{equation}
  where $(i,s)$ is the basis of first uncoupled dot, and $(m,\tau)$ is the basis of two coupled dots 
without interaction; $\psi_{m\tau}(is)$ is the wave function of electron in the first dot in $(m,\tau)$
basis.

  In full analogy with derivations in Ref.[\cite{Zelyak_Murthy_2007}] for GOE to GSE crossover one gets
the following expression for four wave function correlator

\begin{equation}
 \langle \psi_{n\tau_n}(\alpha) \psi^{\ast}_{n\tau_n}(\beta) 
         \psi_{m\tau_m}(\gamma) \psi^{\ast}_{m\tau_m}(\nu) \rangle
 = \frac{\delta^2}{8\pi^2} \big[ \delta_{\alpha\nu} \delta_{\beta\gamma} \Re[D_1] 
   + \delta_{\alpha\gamma} \delta_{\beta\nu} \Re[C_1] \big]
  \label{eq:4w.f_average}
\end{equation}
  where $\Re[D_1]$ and $\Re[C_1]$ are real parts of diffuson and Cooperon contributions to the 
two particle Green's function. Mean level spacing $\delta$ of coupled dot system is 
$\delta = \delta_{1}/2$.

  Comparing indices in Eqs. \eqref{eq:MM_average} and \eqref{eq:4w.f_average} after summation 
over $i$ and $i^{'}$ it is easy to see that the diffuson contribution is $N_1^2$ times 
larger than that of Cooperon. Therefore, the Cooperon contribution is ignored in large-N 
approximation.

  Substituting \eqref{eq:4w.f_average} into \eqref{eq:MM_average} and using 
results of Appendix \ref{apnx:A} for $\Re[D_1]$ one obtains 

\begin{equation}
 \langle M^a_{m\tau,m^{'}\tau^{'}} M^b_{m^{'}\tau^{'},m\tau} \rangle
 = \delta_{ab}\ \frac{\delta_1}{32\pi}\; \frac{E_U}{E_2^2 - E_1^2}
   \bigg[ \frac{E^{2}_{X_2} + E_{X_2}E_U - E^{2}_{1}}{\omega^2 + E_1^2}  
        - \frac{E_{X_2}^2 + E_{X_2}E_U - E_2^2}{\omega^2 + E_2^2}  \bigg]
  \label{eq:MM_averaged}
\end{equation}
   Here $E_U$ and $E_{X_2}$ are the crossover energy scales 
defined in Appendix \ref{apnx:A}. Energy scales $E_{1,2}$ are equal to
$E_{1,2} = a_{1,2}E_{X_2}$, where

\begin{equation*}
 a^2_{1,2} = \frac{4b^2+2b+1 \pm \sqrt{(4b^2+2b+1)^2-4b^2}}{2}
\end{equation*}
   with $b$ defined as $b=E_U/E_{X_2}$.


\section{Spin-spin correlator }\label{apnx:C}

  The spin-spin correlator (spin Green's function) is defined by:

\begin{equation}
 \langle S^a(t) S^b(t^{'}) \rangle = Z^{-1} \int \mathcal{D}{\bf h} \mathcal{D} \bar{\eta}
  \mathcal{D}\eta\, S^a(t) S^b(t^{'}) e^{-S}
\end{equation}
  where $S^a(t)$ is a component of total spin of the system. Let's split the action $S$ 
  (defined by Eq. \eqref{eq:Lagrangian}) in two parts $S = S_1 + S_2$. Here $S_1$ is the part of
the action containing spin $S^a$, $S_1 = -\int dt h^a S^a(t)$, and $S_2 = S - S_1$ contains 
everything else.

  Then spin-spin correlator can be written as

\begin{equation}
 \langle S^a(t) S^b(t^{'}) \rangle = Z^{-1} \int\mathcal{D}{\bf h} \mathcal{D} \bar{\eta}
  \mathcal{D}\eta\, \big[ \frac{\partial}{\partial h^a(t)} \frac{\partial}{\partial h^b(t^{'})} 
    e^{-S_1} \big] e^{-S_2}.
  \label{eq:SS_correl_1}
\end{equation}

  Integration by parts in Eq. \eqref{eq:SS_correl_1} transfers functional derivative  
on $\exp(-S_2)$ term. Performing differentiation one obtains relation

\begin{equation}
 \langle S^a(t) S^b(t^{'}) \rangle = - \frac{\delta(t-t^{'})}{2J}\,\delta_{ab} + 
  \frac{1}{4J^2}\, \langle h^a(t) h^b(t^{'})  \rangle
  \label{eq:SS_correl_2}
\end{equation}

 In Fourier space relation \eqref{eq:SS_correl_2} reads

\begin{equation}
 \langle S^a(i\omega_n) S^b(-i\omega_n) \rangle = - \frac{\delta_{ab}}{2J} + 
  \frac{1}{4J^2}\, \langle h^a(i\omega_n) h^b(-i\omega_n) \rangle
\end{equation}


\section{ Quasiparticle decay rate}\label{apnx:D}

  The interacting self-energy of electron \includegraphics[width=1.0cm]{fig6.pdf} in Matsubara formalism 
is evaluated to

\begin{multline}
  \Sigma^{(1)} = -\beta^{-1} \sum_{\omega_n, \gamma} \mathcal{G}_0(ip_n - i\omega_n) \mathcal{D}_0(i\omega_n)
  \langle\vert M_{\alpha\gamma} \vert^2 \rangle \\
  = -\beta^{-1} \sum_{\gamma} \langle\vert M_{\alpha\gamma} \vert^2 \rangle \sum_{\omega_n}
  \iint_{-\infty}^{\infty} \frac{d\omega^{'} d\omega^{''}}{(2\pi)^2}
  \frac{A_{\gamma}(\omega^{'})}{ip_n - i\omega_n - \omega^{'}}  \frac{H(\omega^{''})}{i\omega_n - \omega^{''}},
\end{multline}
  where $\mathcal{G}_0$ and $\mathcal{D}_0$ are non-interacting Green's functions for electron and bosonic
excitation and $A(\omega^{'})$ and $H(\omega^{''})$ are their spectral representations; 
$\omega_n = 2n\pi/\beta$  and $p_n = (2n+1)\pi/\beta$ are even and odd Matsubara frequencies.

  After summation over $\omega_n$ imaginary part of self energy reads

\begin{equation}
  \Im[\Sigma^{(1)}] = -\frac{1}{2\delta_1} \int_0^{\omega} \langle\vert M_{\alpha\omega^{'}} \vert^2 \rangle
  H(\omega - \omega^{'}) d\omega^{'}.
\end{equation}
  Here we employed the non-interacting expression for electron spectral function 
$A_{\gamma}(\omega^{'}) = 2\pi\delta(\omega^{'} - E_{\gamma})$ assuming small broadening of elergy levels.
Spectral function $H(\omega)$ for bosonic excitations is
\begin{equation}
  H(\omega) = -2\Im\Big[\langle h^a(\omega) h^a(-\omega) \rangle\Big] =
  8JE_U \frac{\omega}{\omega^2 + E_{QCX}^2}.
  \label{eq:H_spectral}
\end{equation}

  Using Eqs. \eqref{eq:MM_averaged} and \eqref{eq:H_spectral} we finally obtain the decay 
rate near the pole

\begin{multline}
  \Im[\Sigma^{(1)}] = -\Gamma/2 = \delta_{ab} \frac{J}{16\pi}\frac{E_U^2}{E_2^2 - E_1^2}
  \bigg[ \frac{E_{X_2}^2+E_{X_2}E_U - E_1^2}{E_1^2 - E_{QCX}^2}\; \ln \frac{E_{QCX}^2(\omega^2 + E_1^2)}{E_1^2
  (\omega^2+E_{QCX}^2)} \\[3mm]
      - \frac{E_{X_2}^2+E_{X_2}E_U - E_2^2}{E_2^2 - E_{QCX}^2}\; \ln \frac{E_{QCX}^2(\omega^2 + E_2^2)}{E_2^2
  (\omega^2+E_{QCX}^2)}  \bigg].
\end{multline}




\bibliographystyle{apsrev}
\bibliography{paper}


\end{document}